\documentclass[11pt]{article}
\usepackage{amssymb}
\usepackage[dvips]{lscape,graphicx}

\newcommand{\bc}{\begin{center}}
\newcommand{\ec}{\end{center}}
\newcommand{\bd}{\begin{displaymath}}
\newcommand{\ed}{\end{displaymath}}
\newcommand{\be}{\begin{equation}}
\newcommand{\ee}{\end{equation}}
\newcommand{\ba}{\begin{array}}
\newcommand{\ea}{\end{array}}
\newcommand{\bt}{\begin{tabular}}
\newcommand{\et}{\end{tabular}}

\newcommand{\ds}{\displaystyle}

\begin{document}

\hyphenation{MSSM NMSSM Z-boson }

\title{Approximate solutions for the Higgs masses and couplings in the
NMSSM}

\author{\underline{R.~Nevzorov}${}^{1,2}$, D.J.~Miller${}^{3}$\\[5mm] 
\itshape{${}^{1}$ School of Physics and Astronomy, University of Southampton, UK} \\[0mm]                          
\itshape{${}^{2}$ Theory Department, ITEP, Moscow, Russia} \\[0mm]   
\itshape{${}^{3}$ Department of Physics and Astronomy, University of Glasgow, UK}}

\date{}

\maketitle

\begin{abstract}
\noindent
We find the approximate solutions for the Higgs masses and couplings in the 
NMSSM with exact and softly broken PQ--symmetry. The obtained solutions indicate 
that there exists a mass hierarchy in the Higgs spectrum which is caused by 
the stability of the physical vacuum.  
\end{abstract}

\section{Introduction}

The minimal SUSY version of the Standard Model (SM) stabilizing the mass hierarchy
does not provide any explanation for its origin.Indeed the Minimal Supersymmetric 
Standard Model (MSSM) being incorporated in the supergravity theories leads to the
$\mbox{$\mu$-problem}$. Within supergravity models the full superpotential
is usually represented as an expansion in powers of observable superfields $\hat{C}_{\alpha}$ 
\be
W=\hat{W}_0(h_m)+\mu(h_m)(\hat{H}_1 \epsilon \hat{H}_2)+
h_{\alpha\beta\gamma}\hat{C}_{\alpha}\hat{C}_{\beta}\hat{C}_{\gamma}+\dots~,
\label{1}
\ee
where $h_m$ and $\hat{W}_0(h_m)$ are the ``hidden'' sector fields and
its superpotential respectively.  The ``hidden'' sector fields acquire
vacuum expectation values of the order of Planck scale ($M_{Pl}$)
breaking local supersymmetry and generating a set of soft masses and
couplings in the observable sector. From dimensional considerations
one would naturally expect the parameter $\mu$ to be either zero or
the Planck scale. If $\mu=0$ then the minimum of the Higgs boson
potential occurs for $\langle H_1 \rangle =0$ and down quarks and
charged leptons remain massless.  In the opposite case, when the
values of $\mu\sim M_{Pl}$, there is no spontaneous breakdown of
$SU(2)\times U(1)$ symmetry at all since the Higgs scalars get a huge
positive contribution $\mu^2$ to their squared masses. In order to
provide the correct pattern of electroweak symmetry breaking, $\mu$ is
required to be of the order of the electroweak scale.
 
In the simplest extension of the MSSM, the Next--to--Minimal
Supersymmetric Standard Model (NMSSM) \cite{3,5}, the
superpotential is invariant with respect to the discrete
transformations $\hat{C}_{\alpha}'=e^{2\pi i/3}C_{\alpha}$ of the
$Z_3$ group. The term $\mu (\hat{H}_1 \hat{H}_2)$ does not meet this
requirement. Therefore it is replaced in the superpotential by
\be
W_{H}=\lambda \hat{S}(H_1 \epsilon H_2)+\frac{1}{3}\kappa\hat{S}^3\,,
\label{4}
\ee 
where $\hat{S}$ is an additional superfield which is a singlet
with respect to $SU(2)$ and $U(1)$ gauge transformations. A
spontaneous breakdown of the electroweak symmetry leads to the
emergence of the vacuum expectation value of the extra singlet field
$\langle S \rangle =s/\sqrt{2}$ and an effective $\mu$--term is
generated ($\mu=\lambda s/\sqrt{2}$). The $Z_3$ symmetry of the
superpotential naturally arises in string inspired models, where all
observable fields are massless in the limit of unbroken supersymmetry.

In this paper we investigate the Higgs masses and their couplings to the
Z--boson in the NMSSM using approximate solutions.  In Section 2
we specify the Higgs sector of the model. In section 3 the exact
Peccei--Quinn (PQ) symmetry limit in the NMSSM is studied and
approximate solutions for the Higgs masses and couplings are
obtained. The scenario of soft PQ--symmetry breaking is discussed
in section 4. In section 5 we summarize our results .

\section{NMSSM Higgs sector}

The NMSSM Higgs sector involves two Higgs doublets $H_{1,2}$ and one
singlet field $S$.  The interactions of the extra complex scalar $S$ with
other particles is defined by the superpotential (\ref{4}) that leads
to a Higgs boson potential of the following form:
\be
\ba{c}
V=\ds\frac{g^2}{8}\left(H_1^+\sigma_a H_1+H_2^+\sigma_a H_2\right)^2+
\frac{{g'}^2}{8}\left(|H_1|^2-|H_2|^2\right)^2+\qquad\qquad\\[3mm]
+\lambda^2|S|^2(|H_1|^2+|H_2|^2)+\lambda\kappa\biggl[S^{*2}(H_1\epsilon
H_2)+h.c.\biggr]+\kappa^2|S|^4+\\[3mm]
+\lambda^2|(H_1\epsilon
H_2)|^2+\biggl[\lambda A_{\lambda}S(H_1\epsilon
H_2)+\ds\frac{\kappa}{3}A_{\kappa}S^3+h.c.\biggr]+\\[3mm]
+m_1^2|H_1|^2+m_2^2|H_2|^2+m_S^2|S|^2+\Delta V\, ,
\ea
\label{5}
\ee
where $g$ and $g'$ are $SU(2)$ and $U(1)$ gauge couplings respectively, 
while $\Delta V$ corresponds to the contribution of loop corrections. The couplings $g,\,g',\,
\lambda$ and $\kappa$ do not violate supersymmetry. The set of soft SUSY breaking
parameters includes soft masses $m_1^2,\, m_2^2,\, m_S^2$ and trilinear couplings $A_{\kappa},\, 
A_{\lambda}$. 

At the physical minimum of the potential (\ref{5}) the neutral
components of the Higgs doublets $H_1$ and $H_2$ develop vacuum
expectation values $v_1$ and $v_2$ breaking the electroweak symmetry
down to $U(1)$. Upon the breakdown of $SU(2)\times U(1)$ symmetry
three goldstone modes ($G^{\pm}$ and $G^{0}$) emerge, and are absorbed
by the $W^{\pm}$ and $Z$ bosons.  In the field space basis rotated by an
angle $\beta$ ($\tan\beta=v_2/v_1$) with respect to the initial direction
\be
\ba{ll}
Im\, H_1^0= (P \sin \beta + G^0 \cos \beta)/\sqrt{2},&\quad H_1^-=G^- \cos \beta + H^- \sin \beta\, , \\
Im\, H_2^0= (P \cos \beta - G^0 \sin \beta)/\sqrt{2},&\quad H_2^+=H^+ \cos \beta - G^+ \sin \beta\, , \\ 
Im\, S= P_S/\sqrt{2}&
\ea
\label{10}
\ee
these unphysical degrees of freedom are removed by a gauge transformation 
and the mass terms in the Higgs boson potential 
can be written as follows
\be
V_{mass} = M_{H^{\pm}}^2  H^+ H^- + 
\frac{1}{2} (P \,\, P_S) \tilde{M}^2 
\left( 
\ba{c} 
P \\ 
P_S 
\ea \right)+
\frac{1}{2} (H \,\, h \,\, N) M^2 
\left( 
\ba{c} 
H \\ 
h \\ 
N
\ea \right)\, ,  
\label{11}
\ee   
where we replace the real parts of the neutral components of the Higgs doublets by their superpositions 
$H\, ,h$ so that
\be
\ba{c}
Re \, H_1^0= (h \cos\beta- H \sin\beta+v_1)/\sqrt{2}\,,\qquad Re\, S= (s+N)/\sqrt{2}\,,\\[2mm]
Re \, H_2^0= (h \sin\beta+ H \cos\beta+v_2)/\sqrt{2}\,.
\ea
\label{12}
\ee
 
From the conditions for the extrema $\left(\ds\frac{\partial V}{\partial v_1}=\frac{\partial V}{\partial v_2}=\frac{\partial V}{\partial s}=0\right)$ 
of the Higgs effective potential (\ref{5}) one can express $m_S^2$, $m_1^2$, $m_2^2$ via other fundamental parameters, $\mbox{tg}\beta$ and $s$. 
Substituting the obtained relations for the soft masses in the $2\times 2$ CP-odd mass matrix $\tilde{M}^2_{ij}$ we get:
\be
\ba{rcl}
\tilde{M}_{11}^2=m_A^2=\ds\frac{4\mu^2}{\sin^2 2\beta}\left(x-\frac{\kappa}{2\lambda}\sin2\beta\right)
+\tilde{\Delta}_{11}\, ,\\
~~~\tilde{M}_{22}^2=\ds\frac{\lambda^2 v^2}{2}x+\frac{\lambda\kappa}{2}v^2\sin2\beta-
3\frac{\kappa}{\lambda}A_{\kappa}\mu+\tilde{\Delta}_{22}\, ,\\
\tilde{M}_{12}^2=\tilde{M}_{21}^2=\ds\sqrt{2}\lambda v \mu\left(\frac{x}{\sin 2\beta}
-2\frac{\kappa}{\lambda}\right)+\tilde{\Delta}_{12}\, ,\\
\ea
\label{14}
\ee
where $v=\sqrt{v_1^2+v_2^2}=246\,\mbox{GeV}$, $\mu=\ds\frac{\lambda s}{\sqrt{2}}$, 
$x=\ds\frac{1}{2\mu}\left(A_{\lambda}+2\frac{\kappa}{\lambda}\mu\right)\sin2\beta$ 
and $\tilde{\Delta}_{ij}$ are contributions of the loop corrections to the mass matrix elements. 
The mass matrix (\ref{14}) can be easily diagonalized via a rotation of the fields 
$P$ and $P_S$ by an angle $\theta_A$ ($\tan 2\theta_A=2\tilde{M}^2_{12}/(\tilde{M}^2_{11}-\tilde{M}^2_{22})$)~.  

The charged Higgs fields $H^{\pm}$ are already physical mass eigenstates with
\be
M_{H^{\pm}}^2=m_A^2-\frac{\lambda^2 v^2}{2}+M_W^2+\Delta_{\pm}.
\label{16}
\ee
Here $M_W=\ds\frac{g}{2}v$ is the charged W-boson mass and $\Delta_{\pm}$ includes loop 
corrections to the charged Higgs masses. 

In the rotated basis  $H\, ,h\,, N$ the matrix elements of the $3\times 3$ mass matrix of 
the CP--even Higgs sector can be written as \cite{6}--\cite{7}:
\be
\ba{rcl}
M_{11}^2&=&\ds m_A^2+\left(\frac{\bar{g}^2}{4}-\frac{\lambda^2}{2}\right)v^2
\sin^2 2\beta+\Delta_{11}\, ,\\[0.2cm]
M_{22}^2&=&\ds M_Z^2\cos^2 2\beta+\frac{\lambda^2}{2}v^2\sin^2 2\beta+
\Delta_{22}\, ,\\[0.2cm]
M_{33}^2&=&\ds 4\frac{\kappa^2}{\lambda^2}\mu^2+\frac{\kappa}{\lambda}A_{\kappa}\mu+
\ds\frac{\lambda^2 v^2}{2}x-\frac{\kappa\lambda}{2}v^2\sin2\beta+\Delta_{33}\, ,\\[0.2cm]
M_{12}^2&=&M_{21}^2=\ds \left(\frac{\lambda^2}{4}-\frac{\bar{g}^2}{8}\right)v^2
\sin 4\beta+\Delta_{12}\, ,\\[0.2cm]
M_{13}^2&=&M_{31}^2=-\sqrt{2}\lambda v \mu x\, \cot  2\beta+\Delta_{13}\, ,
\\[0.2cm]
M_{23}^2&=&M_{32}^2=\sqrt{2}\lambda v \mu (1-x)+\Delta_{23}\, ,
\ea
\label{18}
\ee
where $M_Z=\ds\frac{\bar{g}}{2}v$ is the Z--boson mass, $\bar{g}=\sqrt{g^2+g'^2}$, and
$\Delta_{ij}$ can be calculated by differentiating $\Delta V$ \cite{6}. Since the minimal 
eigenvalue of a matrix does not exceed its smallest diagonal element, at least one Higgs scalar in 
the CP--even sector has to be comparatively light: $m_{h_1}\le \sqrt{M_{22}^2}$. At the tree level 
the upper bound on the lightest Higgs mass in the NMSSM was found in \cite{5}. It differs from the 
corresponding theoretical limit in the minimal SUSY model only for moderate values of $\tan\beta$. 
As in the MSSM the loop corrections from the $t$--quark and its superpartners raise the value of 
the upper bound on the lightest Higgs mass resulting in a rather strict restriction on 
$m_{h_1}\le 135\,\mbox{GeV}$ \cite{8}. The Higgs sector of the NMSSM and loop corrections to it
were studied in \cite{9}.  
 
In the field basis $P,\, P_{S},\,H,\,h,\,N$ the trilinear part of the Lagrangian, which is responsible 
for the interaction of the neutral Higgs states with the Z--boson, is simplified:
\be
L_{AZH}=\ds\frac{\bar{g}}{2} M_{Z}Z_{\mu}Z_{\mu}h+\frac{\bar{g}}{2}Z_{\mu}
\biggl[H(\partial_{\mu}P)-(\partial_{\mu}H)P\biggr]~.
\label{13}
\ee
Only one CP-even component, $h$, couples to a pair of Z--bosons while another, $H$, interacts with
pseudoscalar $P$ and $Z$. 
The coupling of $h$ to the Z pair is exactly the same as in the SM. In the Yukawa 
interactions with fermions $h$ also manifests itself as the SM like Higgs boson.

The couplings of the physical Higgs scalars to the Z pair ($g_{ZZi}$, $i=1,2,3$) and to the Higgs pseudoscalars and Z boson
($g_{ZA_1i}$ and $g_{ZA_2i}$) appear due to the mixing of $h, H$ and $P$ with other components of
the CP--odd and CP--even Higgs sectors. Following the traditional notation we define the normalized 
$R$--couplings as: $g_{ZZi}=R_{ZZi}\times\ds\frac{\bar{g}}{2}M_Z$ and 
$g_{ZA_{j}i}=\ds\frac{\bar{g}}{2}R_{ZA_{j}i}$. All relative $R$--couplings vary from zero to unity
and are given by
\be
R_{ZZi}=U^+_{hi}~,~~~R_{ZA_{1}i}=-U^+_{Hi}\sin\theta_A~,~~~
R_{ZA_2i}=U^+_{Hi}\cos\theta_A~,
\label{20}
\ee
where $U_{ij}$ is unitary matrix relating components of the field basis $H,\,h,$ and $N$ to the physical 
CP-even Higgs eigenstates.

\section{Exact Peccei--Quinn symmetry limit} 

First of all let us discuss the NMSSM with $\kappa=0$. 
At the tree level the Higgs masses and couplings in this model depend on four parameters:
$\lambda,\mu,\tan\beta, m_A (\mbox{or}\,\, x)$.
When $\lambda$ is small enough (say $\lambda \le 0.1$) the experimental constraints on the SUSY parameters 
obtained in the minimal SUSY model remain valid in the the NMSSM. If $\tan\beta\le 2.5$ the predominant 
part of the NMSSM parameter space is excluded by unsuccessful Higgs searches. Non-observation of 
charginos at LEPII restricts the effective $\mu$-term from below: $|\mu|\ge 90-100\,\mbox{GeV}$. 
Combining these limits one gets a useful lower bound on $m_A$ at the tree level:
\be
m_A^2\ge 9M_Z^2 x\,.
\label{25}
\ee
Requirement of the validity of perturbation theory up to the high energy scales constrains the
parameter space further. In order to prevent the appearance of Landau pole during the evolution of the Yukawa couplings 
from the electroweak scale to Grand Unification scale ($M_X$) the value of $\lambda$ has to be always smaller than 0.7 .

In the NMSSM with $\kappa=0$ the mass of the lightest pseudoscalar vanishes. This is a manifestation of the 
enlarged $SU(2)\times [U(1)]^2$ global symmetry of the Lagrangian. The extra $U(1)$ (Peccei--Quinn) symmetry is 
spontaneously broken giving rise to a massless Goldstone boson (axion) \cite{10}. The Peccei--Quinn symmetry 
and the axion allow one to avoid the strong CP problem, eliminating the $\theta$--term in QCD \cite{11}. At low 
energies the axion gains a small mass due to mixing with the pion. The mass of orthogonal superposition of $P$ and $P_S$ is
\be
m_{A_2}^2=m_A^2+\frac{\lambda^2 v^2}{2}x+\tilde{\Delta}_{22}\,.
\label{22}
\ee

The lower bound on $m_A$ (\ref{25}) leads to the hierarchical structure of the CP--even Higgs mass matrix.
It can be written as 
\be 
M^2= 
\left(
\ba{cc}
A & \varepsilon C^{\dagger} \\[3mm]
\varepsilon C  & \varepsilon^2 B
\ea 
\right), 
\label{23}
\ee
where $\varepsilon<<1$. Indeed the top--left entry ($M_{11}^2=A$) of the corresponding $3\times 3$ mass matrix (\ref{18}) 
is the largest one in the dominant part of parameter space. It is proportional to $ m_A^2$ while 
$M_{12}^2\sim M_{22}^2\sim M_{33}^2 \sim M_Z^2$ and $M_{13}^2\sim m_A M_Z$.
Therefore the ratio $M_Z/m_A$ plays the role of a small parameter $\varepsilon$. 

The CP--even Higgs mass matrix can be reduced to block diagonal form:
\be 
V M^2 V^{\dagger}\simeq
\left(
\ba{ccc}
\ds M_{11}^2+\frac{M_{13}^4}{M_{11}^2} & O(\varepsilon^3) & O(\varepsilon^3)\\[3mm]
O(\varepsilon^3) & M_{22}^2 & \ds M_{23}^2-\frac{M_{13}^2M_{12}^2}{M_{11}^2}\\[3mm]
O(\varepsilon^3) & \ds M_{23}^2-\frac{M_{13}^2M_{12}^2}{M_{11}^2}& \ds M_{33}^2-\frac{M_{13}^4}{M_{11}^2} 
\ea 
\right) 
\label{24}
\ee
by virtue of unitary transformation 
\be 
V= 
\left(
\ba{cc}
\ds 1-\frac{\varepsilon^2}{2}\Gamma^{\dagger}\Gamma & \varepsilon \Gamma^{\dagger} \\[3mm]
-\varepsilon \Gamma  &\ds 1-\frac{\varepsilon^2}{2}\Gamma\Gamma^{\dagger}
\ea 
\right),\qquad \Gamma=CA^{-1}\,.
\label{241}
\ee
The obtained matrix (\ref{24}) is now easily diagonalized via a rotation by an angle $\theta$ of the 
two lowest states
\be 
R=
\left(
\ba{ccc}
1 & 0 & 0\\
0 & \cos\theta & \sin\theta\\
0 & -\sin\theta & \cos\theta 
\ea 
\right), \qquad tg\, 2\theta=\frac{\ds 2\left(M_{23}^2-\frac{M_{13}^2M_{12}^2}{M_{11}^2}\right)}
{\ds M_{22}^2-M_{33}^2+\frac{M_{13}^4}{M_{11}^2}} .
\label{243}
\ee
As a result we find the approximate formulae for the masses of the CP--even Higgs bosons
\be
\ba{rcl}
m_{h_3}^2&=&\ds M_{11}^2+\frac{M_{13}^4}{M_{11}^2}\, ,\\[3mm]
m_{h_2,h_1}^2&=&\ds\frac{1}{2}\left(M_{22}^2+M_{33}^2-\frac{M_{13}^4}{M_{11}^2} \right. \\[0mm]
&&\left. \ds\pm\sqrt{\left(M_{22}^2-M_{33}^2+\frac{M_{13}^4}{M_{11}^2}\right)^2+
4\left(M_{23}^2-\frac{M_{13}^2M_{12}^2}{M_{11}^2}\right)^2}\right).
\ea
\label{244}
\ee
Also using the explicit form of the unitary matrix $U^{\dagger}\approx V^{\dagger}R^{\dagger}$, that
links $H, h$ and $N$ to the mass eigenstates, the approximate expressions for the couplings of the
lightest Higgs particles to the Z--boson can be established:
\be
\ba{l}
R_{ZZ2}\approx\cos\theta~,~~~~~~R_{ZZ1}\approx-\sin\theta~,\\[3mm]
R_{ZA_12}\approx\left(\ds\frac{M_{12}^2}{M_{11}^2}\cos\theta+
\frac{M_{13}^2}{M_{11}^2}\sin\theta\right)\sin\theta_A~,\\[3mm]
R_{ZA_11}\approx\left(\ds\frac{M_{13}^2}{M_{11}^2}\cos\theta-
\frac{M_{12}^2}{M_{11}^2}\sin\theta\right)\sin\theta_A~,
\ea
\label{42}
\ee

The approximate solutions for the CP-even Higgs boson masses and couplings shed light on their behaviour 
as the NMSSM parameters are varied. As evident from Eq.(\ref{244}), at large values of $\tan\beta$ or $\mu$ 
the mass--squared of the lightest Higgs scalar tends to be negative because $M_{23}^2$ becomes large
while bottom--right entry of 
the matrix (\ref{24}) goes to zero. Due to the vacuum stability requirement, 
which implies the positivity of the mass--squared of all Higgs particles,
the auxiliary variable $x$ is localized near unity. At the tree level we get
\be
1-\Delta < x < 1+\Delta\, ,\qquad
\Delta\approx\frac{\sqrt{\ds M_Z^2\cos^22\beta+\frac{\lambda^2 v^2}{2}\sin^22\beta}}{m^0_A}\, ,
\label{31}
\ee
where $m_A^0=2\mu/\sin 2\beta$.
The allowed range of the auxiliary variable $x$ is quite narrow (see \cite{7}). 
According to the definition of $m_A$ (\ref{14}) the tight bounds on $x$  
enforce $m_A$ to be confined in the vicinity of $\mu\, tg\beta$ which is considerably larger than the 
Z-boson mass. As a result the masses of the charged Higgs boson, heaviest CP-odd and 
CP-even Higgs states are rather close to $m_A$. At the tree level the theoretical bounds on the 
masses of the lightest Higgs scalars are
\be
\ba{l}
m_{h_1}^2\le \ds\frac{\lambda^2 v^2}{2}x\sin^2 2\beta\,,\\[3mm]
m_{h_2}^2\ge \ds M_Z^2\cos^22\beta+\frac{\lambda^2}{2}v^2\sin^22\beta\, ,\\[3mm]
m_{h_2}^2 \ds\le M_Z^2\cos^22\beta+\frac{\lambda^2}{2}v^2(1+x)\sin^22\beta\, .
\ea
\label{32}
\ee
The masses of $h_2$ and $h_1$ are set by the Z-boson mass and $\lambda v$ respectively so that 
$m_{h_1}, m_{h_2}\ll m_{A}$ in the allowed range of the parameter space. 

\section{Soft breaking of the PQ-symmetry}

Unfortunately searches for massless pseudoscalar and light scalar particles exclude any choice of 
the parameters in the NMSSM with $\kappa=0$, unless one allows $\lambda$ to become very small \cite{12}. 
In order to get a reliable pattern for the Higgs masses and 
couplings the Peccei--Quinn symmetry must be broken. Recently different origins of extra U(1) symmetry breaking 
were discussed \cite{13}-\cite{131}. Here we assume that the violation 
of the Peccei--Quinn symmetry is caused by non--zero value of $\kappa$. As follows from the explicit form of the 
mass matrices (\ref{14}), (\ref{16}) and (\ref{18}) in this case the Higgs spectrum depends on six parameters
at the tree level: $\lambda, \kappa, \mu, tg\beta, A_{\kappa}$ and $m_A$ (or $x$). We restrict our consideration
by small values of $\kappa$ when the PQ--symmetry is only slightly broken. To be precise we consider such values
of $\kappa$ that do not change much the vacuum energy density. the last requirement places a strong bound on $\kappa$ when 
$\lambda$ goes to zero:
\be
\kappa<\lambda^2\,.
\label{34}
\ee 
If $\kappa \gg \lambda^2$ then the terms $\kappa^2|S|^4$ and $\ds\frac{\kappa}{3}A_{\kappa}S^3$ in the Higgs effective 
potential (\ref{5}) becomes much larger $|\mu|^4\sim M_Z^4$ increasing the absolute value of the vacuum energy
density significantly. A small ratio $\kappa/\lambda$ may naturally arise from the renormalization group flow of 
$\lambda$ and $\kappa$ from $M_X$ to $M_Z$ \cite{7,14}.

The soft breaking of the PQ--symmetry does not lead to the realignment of the Higgs spectrum preserving its 
mass hierarchy. Still $M_{11}^2$ is the largest matrix element of the CP-even Higgs mass matrix 
in the admissible part of the NMSSM parameter space. Therefore the approximate formulae (\ref{244})--(\ref{42})
obtained in the previous section remain valid in the considered limit. It is easy to see that
the lightest CP--even Higgs states respect a sum rule
\be
m_{h_1}^2+m_{h_2}^2=M_{22}^2+M_{33}^2-\frac{M_{13}^4}{M_{11}^2}~.
\label{37}
\ee 
The right--hand side of Eq.(\ref{37}) is almost insensitive to the choice of $m_A$ and rather weakly varies with
changing $tg\beta$. As a result of the sum rule (\ref{37}) the second lightest Higgs scalar mass is maximized as $m_{h_1}$ 
goes to zero, and vice versa the lightest Higgs scalar gets maximal mass when $m_{h_2}$ attains minimum 
(see also Fig.1). According to Eq.(\ref{244}) the mass of the lightest CP--even Higgs 
boson vary within the limits: 
\be
\ba{rcccl}
0&\le & m_{h_1}^2 &\le & min\left\{\ds M_{22}^2\, ,\,M_{33}^2-\frac{M_{13}^4}{M_{11}^2}\right\}
\ea
\label{38}
\ee

The mass matrix of the CP-odd Higgs sector also exhibits the hierarchical structure. Indeed the  
entry $\tilde{M}^2_{11}$ is determined by $m_A^2$ whereas the off-diagonal element of the matrix (\ref{14}) is of the order
of $\lambda v \cdot m_A$. Since the ratio $\kappa/\lambda$ is small, the other diagonal entry
$\tilde{M}^2_{22} \ll m_A^2$. this again permits one to seek the eigenvalues of the matrix (\ref{14}) as an expansion in 
powers of $\lambda v/m_A$. The perturbation theory being applied for its diagonalization 
results in concise expressions for the squared masses of the Higgs pseudoscalars
\be
m^2_{A_1}\approx\tilde{M}^2_{22}+\frac{\tilde{M}^4_{12}}{\tilde{M}^2_{11}}\,,\qquad\qquad
m^2_{A_2}\approx\tilde{M}^2_{11}-\frac{\tilde{M}^4_{12}}{\tilde{M}^2_{11}}\,.
\label{39}
\ee 
Because the PQ--symmetry is now broken the lightest CP--odd Higgs boson also gains non--zero mass.
In compliance with Eq.(\ref{42}) the couplings of $h_1$ and $h_2$ to a Z--pair obey the following sum rule:  
\be
R_{ZZ1}^2+R_{ZZ2}^2\simeq 1\, ,
\label{33}
\ee
while $R_{ZA_11}$ and $R_{ZA_12}$ are suppressed by a factor $(\lambda^2 v^2/m_A^2)$.

At the tree level and large values of $tg\beta$ ($tg\beta \gtrsim 10$) the approximate expressions
(\ref{244}) and (\ref{39}) describing the masses of the Higgs scalar and pseudoscalar particles are simplified
\be
\ba{rcl}
m^2_{h_3}&=&m_{A_2}^2=\ds m_A^2+\frac{\lambda^2v^2}{2}x\, ,\qquad\qquad 
m^2_{A_1}=-3\frac{\kappa}{\lambda}A_{\kappa}\mu\, ,\\[3mm]
m^2_{h_2,\, h_1}&=&\ds\frac{1}{2}\left[M_Z^2+4\frac{\kappa^2}{\lambda^2}\mu^2+\frac{\kappa}{\lambda}A_{\kappa}\mu
\pm\right.\\[3mm]
&&\left.\pm\sqrt{\left(\ds M_Z^2-4\frac{\kappa^2}{\lambda^2}\mu^2-\frac{\kappa}{\lambda}A_{\kappa}\mu\right)^2+
8\lambda^2v^2\mu^2(1-x)^2}\right]\, 
\ea
\label{40}
\ee
making their analysis more transparent. Again the positivity of the mass--squared of the
lightest Higgs scalar restricts the allowed range of $x$
\be
1-\left|\frac{\sqrt{2}\kappa M_Z}{\lambda^2 v}\right|<x<1+\left|\frac{\sqrt{2}\kappa M_Z}{\lambda^2 v}\right|\,,
\label{41}
\ee
if the PQ--symmetry is only slightly broken. When $\kappa\ll \lambda^2$ the admissible interval of the auxiliary 
variable $x$ shrinks drastically establishing a very stringent bound on the value of $m_A$ and strong correlation 
between $m_A$, $\mu$ and $tg\, \beta$ (see \cite{7}, \cite{131}) that constrains the masses of the heavy Higgs bosons 
$m_{h_3}\approx m_{H^{\pm}}\approx m_{A_2}$ in the vicinity of $\mu\, tg\beta$. 

The results of the numerical studies of the Higgs boson masses and their couplings including leading one--loop 
corrections from the top and stop loops are given in Fig.1--3. As a representative example we fix the Yukawa couplings
at the Grand Unification scale so that $\lambda(M_X)=\kappa(M_X)=2h_t(M_X)=1.6$, that corresponds to $tg\beta\ge 3$,  
$\lambda(M_t)\simeq 0.6$ and $\kappa(M_t)\simeq 0.36$ at the electroweak scale.
We set $\mu=150\,\mbox{GeV}$ which is quite close to the current limit on $\mu$ in the MSSM.
The parameter $A_{\kappa}$ occurs in the right--hand side of the sum rule (\ref{37}) and in the mass of the lightest 
pseudoscalar $m_{A_1}^2$ with opposite sign. As a consequence whereas $m_{h_1}^2$ rises over changing $A_{\kappa}$, $m_{A_1}^2$ 
diminishes and vice versa. Too large positive and negative values of $A_{\kappa}$ pull the mass-squared 
of either lightest scalar or pseudoscalar below zero destabilizing the vacuum that restricts $A_{\kappa}$
from below and above. To represent the results of our numerical analysis the parameter $A_{\kappa}$ is taken to be 
near the center of the admissible interval $A_{\kappa}\simeq 135\,\mbox{GeV}$. 

\vspace{2mm}
\begin{center}
{\hspace*{-0mm}\includegraphics[totalheight=60mm,keepaspectratio=true]{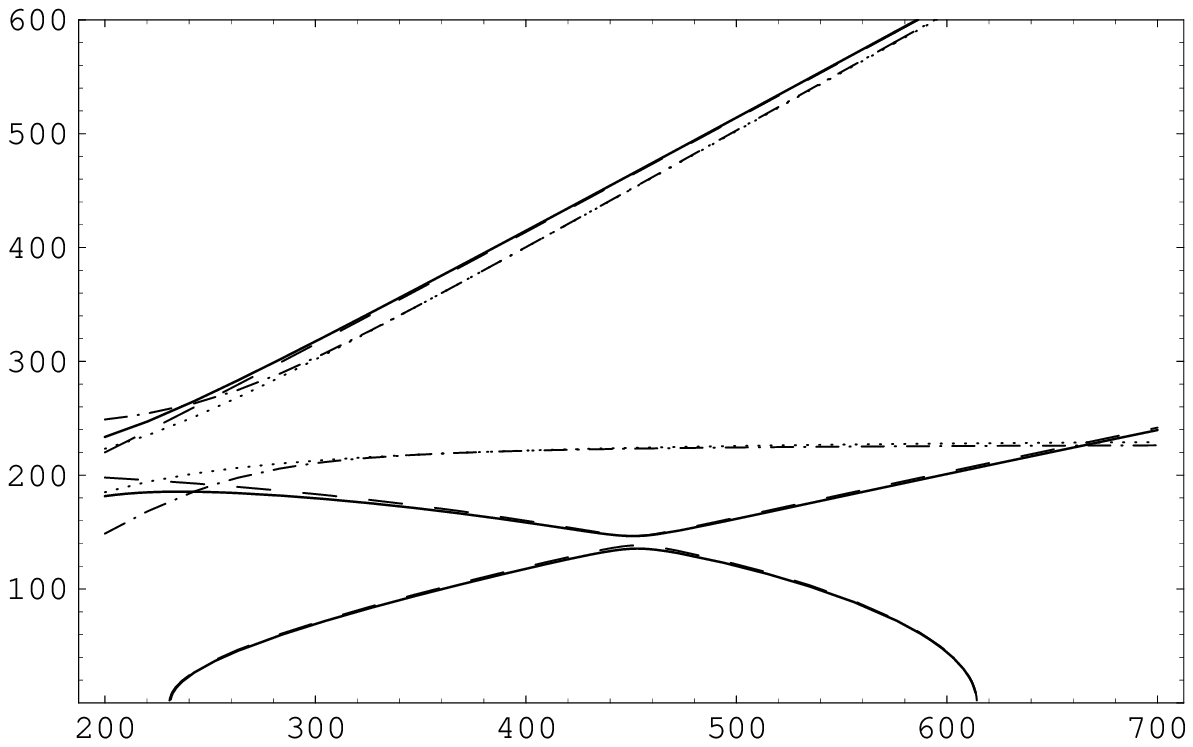}}\\
{\large $m_{A}$ (GeV)}\\[2mm]
\end{center}
{Figure 1.{\it The dependence of the neutral Higgs boson masses on $m_A$ for 
$\lambda=0.6$, $\kappa=0.36$, $\mu=150$\,GeV, $tg \beta=3$ and $A_{\kappa}=135\,\mbox{GeV}$.
The one--loop Higgs masses of scalars (solid curve) and pseudoscalars (dashed--dotted curve) are 
confronted with the approximate solutions (dashed and dotted curves). Masses are in GeV.}}\\

In Fig.1--3 the masses of the neutral Higgs particles and their
couplings to Z are examined as a function of $m_A$.  From the
restrictions (\ref{41}) on the parameter $x$ and numerical results
presented in Fig.1 it is evident that the requirement of the stability
of the physical vacuum and the experimental constraints on $\mu$ and
$tg\beta$ rules out low values of $m_A$, maintaining mass hierarchy
whilst $\kappa\le\lambda^2$. The lightest Higgs scalar and
pseudoscalar can be heavy enough to escape their production at
LEP. Moreover as one can see from Fig.2 the lightest Higgs scalar can
be predominantly a singlet field, making its detection more difficult
than in the SM or MSSM.  The lightest Higgs pseudoscalar is also
singlet dominated, making its observation at future colliders quite
problematic; the coupling of the lightest CP--even Higgs boson to a
CP--odd Higgs bosons and a Z is always strongly suppressed (see Fig.3)
according to (\ref{42}). The hierarchical structure of the mass
matrices ensures that the heaviest CP-even and CP-odd Higgs bosons are
predominantly composed of $H$ and $P$. As a result the coupling 
$R_{ZA_23}$ is rather close to unity while $R_{ZZ3}$ is
almost negligible.  In Fig.1--3 the approximate solutions
(\ref{244})--(\ref{42}) are also given. They work remarkably well.
%
\vspace{1mm}
\begin{center}
{\hspace*{-0mm}\includegraphics[height=60mm,keepaspectratio=true]{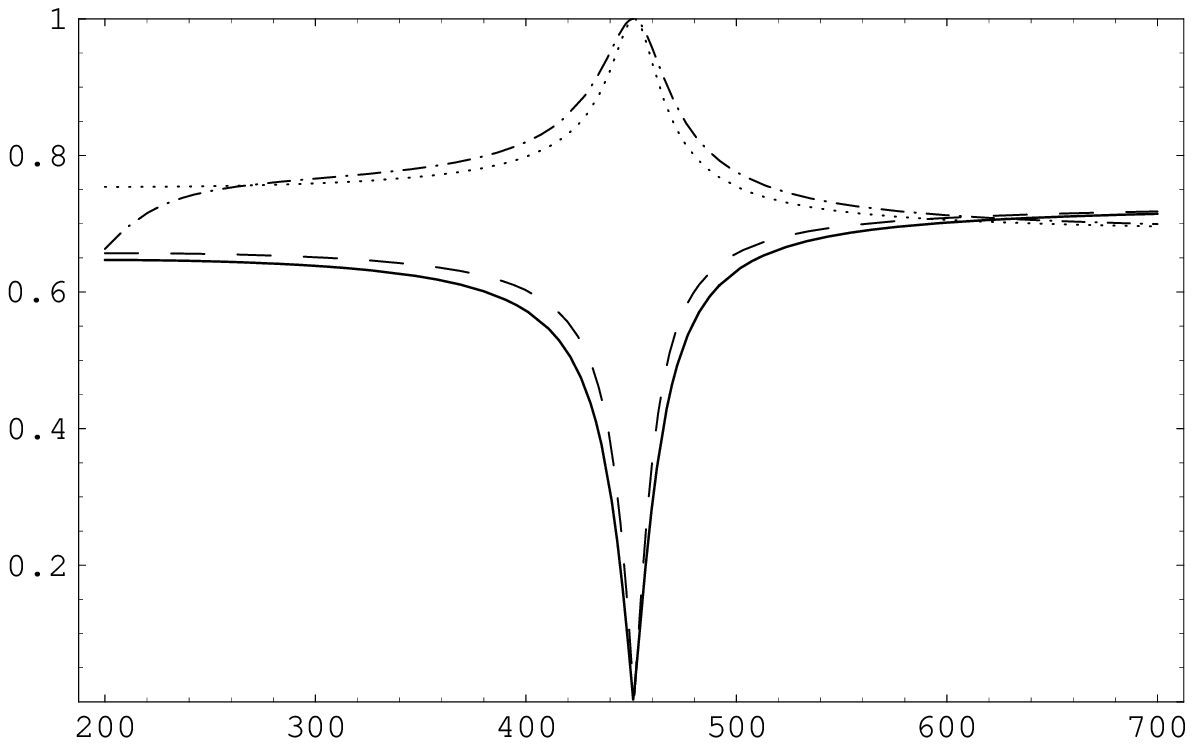}}\\
{\large $m_{A}$ (GeV)}\\[2mm]
\end{center}
{Figure 2.{\it The absolute values of $R_{ZZ1}$ and $R_{ZZ2}$,
plotted as a function of $m_A$ for the same values of $\lambda$, $\kappa$, $\mu$, $\tan\beta$ and $A_{\kappa}$
as in Fig.1. Solid and dashed--dotted curves reproduce the dependence of $R_{ZZ1}$ and 
$R_{ZZ2}$ on $m_A$ while dashed and dotted curves represent their approximate solutions.}}\\

\section{Conclusions}

In the present article we have obtained the approximate solutions for the Higgs masses and couplings in the
NMSSM with exact and softly broken PQ-symmetry which describe the numerical solutions with high accuracy. 
The approximate formulae (\ref{244})--(\ref{42}) provide nice insight into mass hierarchies
in the considered model. The vacuum stability requirements and LEP restrictions on the NMSSM parameters 
leads to the splitting in the spectrum of the Higgs bosons. When $\kappa=0$ or $\kappa\le\lambda^2$
the charged Higgs states, the heaviest scalar and pseudoscalar are nearly degenerate around $m_A\sim\mu\, tg\beta$.
The masses of new scalar and pseudoscalar states, which are predominantly singlet fields, are governed by the
combination of parameters $\ds\frac{\kappa}{\lambda}\,\mu$. In the NMSSM with exact and softly broken PQ--symmetry
they are considerably lighter than the heaviest Higgs states. Decreasing $\kappa$ pushes their masses
down so that they can be even the lightest particles in the Higgs boson spectrum. The SM like Higgs boson has a mass around $130\,\mbox{GeV}$. We have established useful sum rules for the masses of the lightest Higgs scalars and 
their couplings to a Z pair. Also we found that the couplings of the lightest CP--even Higgs states to the lightest 
pseudoscalar 
and Z--boson are suppressed. Observing two light scalar and one pseudoscalar Higgs particles but no charged Higgs 
boson, at future colliders would present an opportunity to distinguish the NMSSM with softly broken PQ--symmetry 
from the MSSM even if the heavy states are inaccessible.

\begin{center}
{\hspace*{-0mm}\includegraphics[height=60mm,keepaspectratio=true]{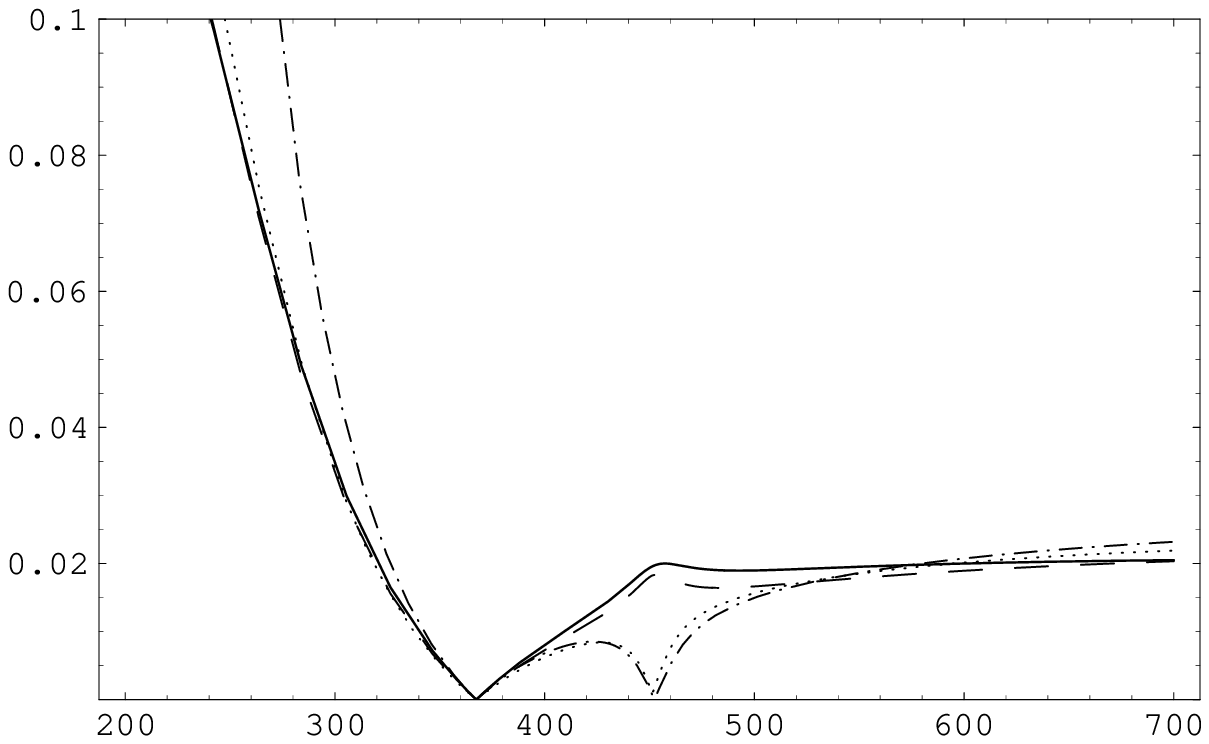}}\\
{\large $m_{A}$ (GeV)}\\[2mm]
\end{center}
{Figure 3.{\it The absolute values of $R_{ZA_11}$ and $R_{ZA_12}$ as a function of $m_A$.
The parameters $\lambda$, $\kappa$, $\mu$, $\tan\beta$ and $A_{\kappa}$ are taken to be the same as in 
Fig.1. Solid and dashed--dotted curves correspond to $R_{ZA_11}$ and $R_{ZA_12}$ while dashed and dotted curves 
represent their approximate solutions.}}

\section*{Acknowledgements}
The authors would like to thank P.M.~Zerwas for his continual support and encouragement.
RN is grateful to C.D.~Froggatt, E.I.~Guendelman, N.S.~Mankoc-Borstnik and H.B.~Nielsen for stimulating questions and 
comments, and S.F.~King, S.~Moretti, A.~Pilaftsis, M.~Sher and M.I.~Vysotsky for fruitful discussions and helpful remarks.
The work of RN was supported by the Russian Foundation for Basic Research (projects 00-15-96562 and 02-02-17379) 
and by a Grant of President of Russia for young scientists (MK--3702.2004.2).


\begin{thebibliography}{99}

\bibitem{3}
P.Fayet, Nucl.Phys. B 90 (1975) 104;
H.P.Nilles, M.Srednicki, D.Wyler, Phys.Lett.B 120 (1983) 346;
J.M.Frere, D.R.T.Jones, S.Raby, Nucl.Phys.B 222 (1983) 11;
J.P.Derendinger, C.A.Savoy, Nucl.Phys.B 237 (1984) 307.
M.I.Vysotsky, K.A.Ter-Martirosian, Sov.Phys.JETP 63 (1986) 489;
J.Ellis, J.F.Gunion, H.Haber, L.Roszkowski, F.Zwirner, Phys.Rev.D 39 
(1989) 844.
\bibitem{5}
L.Durand, J.L.Lopez, Phys.Lett.B 217 (1989) 463;
L.Drees, Int.J.Mod.Phys.A 4 (1989) 3635.
\bibitem{6}
P.A.Kovalenko, R.B.Nevzorov, K.A.Ter--Martirosyan, Phys.Atom.Nucl. 61 (1998) 812.
\bibitem{7}
D.J.Miller, R.Nevzorov, P.M.Zerwas, Nucl.Phys.B 681 (2004) 3.
\bibitem{8}
M.Masip, R.Mu$\tilde{n}$oz--Tapia, A.Pomarol, Phys.Rev.D 57 (1998) 5340; G.K.Yeghian, hep--ph/9904488;
U.Ellwanger, C.Hugonie, Eur.Phys.J.C 25 (2002) 297.
\bibitem{9}
T.Elliott, S.F.King, P.L.White, Phys.Lett.B 314 (1993) 56; U.Ellwanger, Phys.Lett.B 303 (1993) 271;
U.Ellwanger, M.Lindner, Phys.Lett.B 301 (1993) 365; P.N.Pandita, Phys.Lett.B 318 (1993) 338;
P.N.Pandita, Z.Phys.C 59 (1993) 575; T.Elliott, S.F.King, P.L.White, Phys.Rev.D 49 (1994) 2435;
S.F.King, P.L.White, Phys.Rev.D 52 (1995) 4183;
\bibitem{10}
S.Weinberg, Phys.Rev.Lett. 40 (1978) 223; F.Wilczek, Phys.Rev.Lett. 40 (1978) 279.
\bibitem{11}
R.D.Peccei, H.R.Quinn, Phys.Rev.Lett. 38 (1977) 1440; Phys.Rev.D 16 (1977) 1791.
\bibitem{12}
D.J.Miller, R.Nevzorov, hep-ph/0309143.
\bibitem{13}
C.Panagiotakopoulos, K.Tamvakis, Phys.Lett.B 469 (1999) 145;
R.B.Nevzorov, M.A.Trusov,J.Exp.Theor.Phys.91 (2000) 1079;
A.Dedes, C.Hugonie, S.Moretti, K.Tamvakis, Phys.Rev.D 63 (2001) 055009;
R.B.Nevzorov, K.A.Ter-Martirosyan, M.A.Trusov, Phys.Atom.Nucl.65 (2002) 285.
\bibitem{131}
C.Panagiotakopoulos, A.Pilaftsis, Phys.Rev.D 63 (2001) 055003.
\bibitem{14}
R.B.Nevzorov, M.A.Trusov, Phys.Atom.Nucl.64 (2001) 1299.


\end{thebibliography}
\end{document}